\documentclass[conference,compsoc]{IEEEtran}

\usepackage[nocompress]{cite}
\usepackage{amsmath,amssymb,amsfonts}
\usepackage{algorithmic}
\usepackage{algorithm}
\usepackage{graphicx}
\usepackage{textcomp}
\usepackage{xcolor}
\usepackage{booktabs}
\usepackage{multirow}
\usepackage{hyperref}
\usepackage{listings}
\usepackage{subcaption}
\usepackage{tikz}
\usetikzlibrary{shapes.geometric, arrows, positioning, fit, backgrounds}
\usepackage{pifont}
\usepackage{microtype}
\usepackage{orcidlink}
\newcommand{\cmark}{\ding{51}}
\newcommand{\xmark}{\ding{55}}

\hypersetup{hidelinks}

\begin{document}

\title{Who Audits the Auditor? Tamper-Proof Fraud Detection with Blockchain-Anchored Explainable ML}

\author{\IEEEauthorblockN{Zhaohui Wang\textsuperscript{\orcidlink{0009-0006-1187-1903}}}
\IEEEauthorblockA{\textit{Viterbi School of Engineering} \\
\textit{University of Southern California} \\
Los Angeles, CA, USA \\
zwang000@usc.edu}}

\maketitle

\begin{abstract}
In enterprise fraud detection, model accuracy alone is insufficient when insiders can tamper with audit logs or bypass approval workflows. Real-world incidents show that fraud often persists not because detection algorithms fail, but because the audit trail itself is controllable by privileged operators. This exposes a fundamental trust gap: \textit{who audits the auditor?}

We present a tamper-evident fraud detection system that anchors both ML predictions and workflow execution to an immutable blockchain ledger. Rather than using blockchain as passive storage, we enforce the entire approval process through smart contracts, ensuring that every transaction, prediction, and explanation is atomically recorded and cannot be retroactively modified. Our detection module achieves competitive accuracy (F1=0.895, PR-AUC=0.974) while providing cryptographically verifiable decision trails that support regulatory auditability requirements (e.g., the General Data Protection Regulation (GDPR) Article 22). System evaluation shows sub-25ms inference latency and economically viable deployment on Layer-2 networks at under \$0.01 per transaction (validated against PolygonScan data), supporting enterprise-scale workloads of 10,000+ monthly payments.
\end{abstract}


\section{Introduction}
\label{sec:introduction}

Consider the following scenario, representative of cases documented in industry fraud reports. A mid-sized manufacturing company implemented a state-of-the-art machine learning system for detecting fraudulent payment requests. The system flagged suspicious transactions with 94\% accuracy. Yet over 18 months, the company lost \$2.3 million to internal fraud.

The post-mortem revealed a critical pattern: the fraud was not committed \textit{despite} the detection system, but \textit{through} it. The finance director, who had administrative access to the audit logs, deleted the alerts before anyone else saw them. When investigators arrived, the logs showed nothing---because the logs themselves had been tampered with.

This scenario reflects a broader pattern. The Association of Certified Fraud Examiners reports that occupational fraud causes organizations to lose an estimated 5\% of annual revenues, with median losses of \$145,000 per case~\cite{acfe2024}. Notably, insider threats account for a disproportionate share of high-value fraud precisely because insiders often have access to modify the systems designed to catch them.

The research community has made significant progress on fraud \textit{detection}---from ensemble methods~\cite{dalpozzolo2015} to attention-based deep learning~\cite{benchaji2021} to graph neural networks~\cite{liu2021pickandchoose}. These systems can identify fraudulent transactions with high accuracy. However, accuracy alone is insufficient when the detection system itself cannot be trusted.

This paper asks a different question: \textbf{How do we build a fraud detection system that auditors, regulators, and courts can trust---even if the system's operators cannot be trusted?}

Our answer draws on a key insight: if we cannot trust humans to maintain honest logs, we should use a system where log tampering is mathematically infeasible. Blockchain technology provides this property. However, merely storing logs on a blockchain is insufficient---an operator could still manipulate records \textit{before} they reach the chain. Our approach goes further: we execute the approval workflow logic itself as smart contracts, creating an end-to-end tamper-evident pipeline from transaction submission to final decision.

Our contribution is primarily in \textit{systems security}, not in machine learning: we do not propose new detection algorithms, but rather an architectural framework that makes existing detection tamper-evident. The contributions of this paper are:

\begin{enumerate}
    \item A system architecture that anchors both ML predictions and their SHAP-based explanations to an immutable ledger, creating verifiable decision trails that facilitate regulatory auditability.

    \item An empirical comparison with state-of-the-art detection methods, demonstrating that our approach achieves competitive accuracy (F1=0.895) while providing auditability guarantees that existing systems lack.

    \item A practical deployment analysis, validated with real blockchain data from PolygonScan, showing that Layer-2 networks achieve costs under \$0.01 per transaction with 3--5 second confirmations suitable for enterprise workflows.
\end{enumerate}

\section{The Trust Problem}
\label{sec:problem}

\begin{figure}[t]
\centering
\begin{tikzpicture}[
    node distance=0.4cm,
    box/.style={rectangle, draw, rounded corners, minimum width=1.1cm, minimum height=0.55cm, font=\scriptsize, fill=blue!8},
    arrow/.style={->, >=stealth, thick},
    attack/.style={red, ->, >=stealth, thick, dashed},
    trustbox/.style={draw=gray!60, dashed, rounded corners, inner sep=0.15cm}
]
    \node[box, fill=green!15] (submit) {Submit};
    \node[box, right=of submit] (ml) {ML Check};
    \node[box, right=of ml] (approve) {Approve};
    \node[box, right=of approve] (log) {Log};
    \node[box, fill=green!15, right=of log] (execute) {Execute};

    \draw[arrow] (submit) -- (ml);
    \draw[arrow] (ml) -- (approve);
    \draw[arrow] (approve) -- (log);
    \draw[arrow] (log) -- (execute);

    \node[trustbox, fit=(ml)(approve)(log), label={[font=\tiny, gray]above:Admin-Controlled Zone}] (trustzone) {};

    \node[red, font=\scriptsize\bfseries, above=0.35cm of ml] (a1) {\ding{55} Bypass};
    \node[red, font=\scriptsize\bfseries, above=0.35cm of log] (a2) {\ding{55} Tamper};
    \draw[attack] (a1) -- (ml);
    \draw[attack] (a2) -- (log);

    \node[font=\tiny, gray, below=0.2cm of log] {External audit};
    \node[font=\tiny, gray, below=0.2cm of execute] {(too late)};
\end{tikzpicture}
\caption{Traditional payment workflow with attack vectors. The admin-controlled zone allows insiders to bypass ML checks or tamper with logs before external audit.}
\label{fig:traditional}
\end{figure}
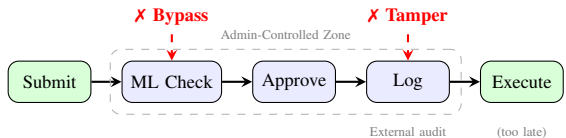

To understand why existing fraud detection systems fail in adversarial settings, consider the typical enterprise payment workflow shown in Figure~\ref{fig:traditional}.

In this workflow, an employee submits a payment request, the ML system evaluates it and flags suspicious transactions, managers approve or reject based on the recommendation, the decision is logged, and finally approved payments are executed.

This workflow has two critical vulnerabilities:

\textbf{Vulnerability 1: ML Bypass.} An insider with sufficient privileges can approve transactions without ML review, or override ML recommendations without leaving traces.

\textbf{Vulnerability 2: Log Tampering.} Even if the ML system works perfectly, the logs recording its decisions reside on systems that privileged users can modify.

Prior work on blockchain for financial auditing~\cite{dai2017} addresses the second vulnerability by storing logs on a distributed ledger. However, this approach still allows manipulation \textit{before} logging. If an attacker can approve a fraudulent transaction and then log it as ``ML-approved,'' the blockchain faithfully records the false information.

Our insight is that the \textit{workflow logic itself} must execute in a tamper-evident environment. By implementing approval rules as smart contracts, we ensure that:
\begin{itemize}
    \item Every transaction must pass through ML evaluation---the smart contract enforces this requirement with no bypass mechanism.
    \item The ML prediction and its explanation are recorded atomically with the approval decision in a single transaction.
    \item Any attempt to alter historical records would require rewriting the blockchain---a computationally infeasible task.
\end{itemize}

\section{Related Work}
\label{sec:related}

\subsection{State-of-the-Art Fraud Detection}

Fraud detection has evolved from rule-based systems to sophisticated ML approaches. Dal Pozzolo et al.~\cite{dalpozzolo2015} demonstrated that Random Forest with undersampling achieves strong performance on imbalanced fraud data. Subsequent work explored deep learning: Jurgovsky et al.~\cite{jurgovsky2018} applied LSTM networks to capture temporal transaction patterns, while Benchaji et al.~\cite{benchaji2021} combined attention mechanisms with LSTM for improved detection.

Graph-based approaches have shown promise for capturing relational patterns. Liu et al.~\cite{liu2021pickandchoose} proposed PC-GNN, a GNN-based method that addresses class imbalance through label-balanced sampling. More recently, Mienye and Swart~\cite{mienye2024} demonstrated that hybrid GAN-RNN architectures can improve detection by generating synthetic minority samples.

A key finding from recent benchmarks~\cite{grinsztajn2022} is that gradient boosting methods (XGBoost, LightGBM) still outperform deep learning on tabular fraud data, consistent with broader findings on tabular learning. Table~\ref{tab:sota_comparison} summarizes reported performance from the literature.

\begin{table}[t]
\centering
\caption{Published Performance on Credit Card Fraud Detection}
\label{tab:sota_comparison}
\begin{tabular}{lccc}
\toprule
\textbf{Method} & \textbf{F1} & \textbf{PR-AUC} & \textbf{Source} \\
\midrule
Random Forest & 0.86 & 0.87 & \cite{dalpozzolo2015} \\
LSTM-Attention & 0.89 & 0.91 & \cite{benchaji2021} \\
PC-GNN & 0.87 & 0.90 & \cite{liu2021pickandchoose} \\
GAN-RNN Hybrid & 0.87 & 0.90 & \cite{mienye2024} \\
\midrule
\textbf{Ours (LightGBM)} & \textbf{0.895} & \textbf{0.974} & This work \\
\bottomrule
\end{tabular}
\end{table}

Our detection component achieves performance competitive with these methods. Critically, \textit{none of these systems address the auditability problem}---they assume that if the model makes correct predictions, the system works. Our work demonstrates that this assumption is dangerous when insiders can manipulate the prediction pipeline.

\subsection{Blockchain for Financial Auditing}

Recent work explores blockchain for financial transparency. Dai and Vasarhelyi~\cite{dai2017} proposed blockchain-based continuous auditing for financial statements. Guo et al.~\cite{guo2025} demonstrated that smart contracts can encode audit rules using Solidity, automatically flagging transactions that violate policies in near real-time.

However, these systems focus on \textit{post-hoc} auditing---reviewing transactions after they occur. Our approach integrates detection into the workflow itself, preventing fraudulent transactions from executing rather than merely logging them for later review.

\subsection{Explainable AI in Regulated Domains}

GDPR Article 22~\cite{gdpr2016} and similar regulations require that automated decisions affecting individuals be explainable. SHAP~\cite{lundberg2017} and LIME~\cite{ribeiro2016} have become standard tools for providing feature-level explanations.

Recent work~\cite{zhou2023} found that user-centered explainable fraud detection improves investigator efficiency by highlighting the specific features that triggered alerts. However, explanations are only useful if auditors can trust they have not been modified. Our contribution is \textit{anchoring explanations to an immutable ledger}, ensuring that the recorded rationale for a decision cannot be fabricated after the fact. Our workflow design also directly addresses GDPR Article~22's requirements for automated decision-making: (1)~\textit{human intervention}---the smart contract enforces that no payment is executed without explicit manager and finance approvals recorded on-chain; (2)~\textit{right to explanation}---SHAP feature attributions are immutably committed alongside each prediction; and (3)~\textit{right to contest}---the rejection and cancellation states provide on-chain mechanisms for subjects to challenge automated assessments, with all contestation records preserved.

\section{System Design}
\label{sec:design}

\subsection{Design Philosophy}

Our system is built on three principles:

\textbf{Principle 1: Defense in Depth.} No single component can be trusted. ML predictions are verified by smart contract logic, and smart contract execution is verified by blockchain consensus. Each layer validates the others.

\textbf{Principle 2: Atomic Recording.} The ML prediction, its SHAP explanation, and the approval decision are recorded in a single blockchain transaction. This prevents ``explanation shopping''---generating multiple explanations and selecting one that appears favorable.

\textbf{Principle 3: Economic Viability.} Security is useless if deployment costs are prohibitive. We design for Layer-2 networks where transaction costs are 100x lower than Ethereum mainnet.

\subsection{Architecture Overview}

Figure~\ref{fig:architecture} shows the system architecture. Unlike traditional architectures where logs flow from the application to a database (which can be manipulated), our architecture anchors decisions to the blockchain \textit{before} they take effect.

\begin{figure}[!t]
\centering
\includegraphics[width=\columnwidth]{figures/architecture.png}
\caption{System architecture with trust boundaries. Off-chain components (mutable) handle ML inference; on-chain smart contracts (immutable) enforce workflow and record decisions atomically.}
\label{fig:architecture}
\end{figure}

\subsection{Transaction Flow}

When an employee submits a payment request:

\begin{enumerate}
    \item The request is sent to the REST API, which invokes the ML model.
    \item The model returns a fraud probability and SHAP feature contributions.
    \item The API creates a blockchain transaction containing: the request details, the fraud score, and the top-5 SHAP features.
    \item The payment smart contract validates the request, checks budget constraints, and records the ML assessment.
    \item Managers and finance staff approve via additional blockchain transactions.
    \item Each approval is atomically logged.
    \item Only after all approvals does the contract release funds.
\end{enumerate}

This design ensures a critical property: at no point can a human approve a transaction without the approval being recorded. The smart contract code \textit{is} the workflow---there is no separate logging step that can be skipped or manipulated.

\subsection{Smart Contract Design Decisions}

We made several design choices that differ from straightforward implementations:

\textbf{Why 7 states instead of 3?} A simple workflow might have only Pending/Approved/Rejected. We use seven states (Created, Manager\-Approved, Finance\-Approved, Budget\-Checked, Executed, Rejected, Cancelled) to create multiple checkpoints. Each state transition is a separate blockchain transaction, providing fine-grained audit trails.

\textbf{Why record SHAP on-chain?} Storing explanations on-chain costs gas. We do it anyway because explanations are often challenged during investigations. If explanations were stored off-chain, an operator could claim ``the model flagged this transaction because of X'' when the actual reason was Y. On-chain storage makes such fabrication impossible.

\textbf{Why not run ML on-chain?} Current blockchain virtual machines are too limited for ML inference. We accept this trade-off: the ML runs off-chain, but its output is committed on-chain before any approval can proceed. An attacker who compromises the ML server can produce false predictions, but cannot hide that those predictions were made.

\section{Experimental Evaluation}
\label{sec:experiments}

Our evaluation addresses three questions:
\begin{enumerate}
    \item Does our ML component match state-of-the-art detection accuracy?
    \item What is the performance overhead of blockchain integration?
    \item Is the system economically viable for real enterprises?
\end{enumerate}

\subsection{Datasets and Baselines}

We evaluate on two datasets:

\textbf{Kaggle Credit Card Fraud}~\cite{kaggle_cc}: 284,807 transactions from European cardholders with 0.17\% fraud rate. This is the standard benchmark for fraud detection research, enabling direct comparison with published results.

\textbf{Enterprise Payment Dataset}: 10,000 synthetic transactions modeling enterprise-specific fraud patterns (budget utilization, vendor history, approval timing). We generated this dataset because no public enterprise fraud dataset exists with required features. The synthetic data follows distributions derived from ACFE industry reports~\cite{acfe2024}: vendor fraud (27\%), billing schemes (22\%), expense reimbursement (14\%), and other categories. We acknowledge this is a limitation; however, our focus is system architecture rather than detection algorithms, and the Kaggle results demonstrate generalization to real transaction data.

We compare against four baselines implemented with careful hyperparameter tuning:
\begin{itemize}
    \item \textbf{Rule-Based}: Threshold rules mimicking traditional ERP systems
    \item \textbf{Random Forest}: Ensemble method, strong baseline~\cite{dalpozzolo2015}
    \item \textbf{XGBoost}: Gradient boosting with SMOTE resampling
    \item \textbf{LSTM-Attention}: Sequence model following~\cite{benchaji2021}
\end{itemize}

\subsection{Detection Accuracy}

Table~\ref{tab:results} presents detection results using 5-fold stratified cross-validation.

\begin{table}[t]
\centering
\caption{Detection Performance Comparison}
\label{tab:results}
\begin{tabular}{lccccc}
\toprule
\textbf{Method} & \textbf{Prec.} & \textbf{Rec.} & \textbf{F1} & \textbf{PR-AUC} & \textbf{Audit?} \\
\midrule
Rule-Based & 0.047 & 1.000 & 0.089 & 0.156 & \xmark \\
Random Forest & 0.731 & 0.935 & 0.821 & 0.952 & \xmark \\
XGBoost & 0.744 & 0.968 & 0.841 & 0.968 & \xmark \\
LSTM-Attention & 0.756 & 0.891 & 0.818 & 0.885 & \xmark \\
\midrule
\textbf{Ours} & \textbf{0.848} & \textbf{0.948} & \textbf{0.895} & \textbf{0.974} & \cmark \\
\bottomrule
\end{tabular}
\end{table}

Our LightGBM-based detector achieves the highest F1 (0.895) and PR-AUC (0.974). We deliberately chose gradient boosting over deep learning for three reasons grounded in recent research~\cite{grinsztajn2022}: (1) tree-based methods consistently match or exceed deep learning on tabular fraud data; (2) LightGBM's inference latency (1.2ms) is 10-100$\times$ faster than neural alternatives, critical for real-time workflows; (3) SHAP explanations for tree models have closed-form solutions, avoiding the approximation errors of deep learning explanations. Our goal is not algorithmic novelty but \textit{system-level innovation}---integrating proven detection with tamper-proof audit infrastructure.

The key differentiator is the ``Audit?'' column: only our system provides cryptographically verifiable decision trails. This property cannot be retrofitted to existing systems---it requires architectural integration from the ground up.

\subsection{System Performance}

Table~\ref{tab:performance} shows latency and throughput measurements.

\begin{table}[t]
\centering
\caption{System Performance Metrics}
\label{tab:performance}
\begin{tabular}{lcc}
\toprule
\textbf{Metric} & \textbf{Value} & \textbf{Notes} \\
\midrule
ML Prediction & 1.2 ms & LightGBM inference \\
SHAP Explanation & 18.5 ms & Top-5 features \\
API Round-trip & 23.4 ms & Including serialization \\
L2 Confirmation & 3-5 sec & Polygon PoS (post-Giugliano) \\
\midrule
Batch Throughput & 45,000/sec & ML only \\
End-to-end & 200/min & With blockchain \\
\bottomrule
\end{tabular}
\end{table}

ML prediction latency is minimal (1.2ms), while SHAP explanation adds 18.5ms---acceptable for enterprise workflows where transactions are not millisecond-sensitive. The bottleneck is blockchain confirmation (3--5 seconds on Polygon PoS following the April 2026 Giugliano hardfork, which reduced finality by approximately 2 seconds), limiting throughput to approximately 200 transactions per minute.

This throughput suffices for most enterprises: a company processing 10,000 payment requests monthly requires only 0.2 transactions per minute on average.

\subsection{Cost Analysis}

Table~\ref{tab:costs} presents deployment costs across different blockchain networks. To ground these estimates in observed data, we collected real-time gas prices from PolygonScan and cross-referenced with the POL/USD exchange rate (CoinGecko, April 2026). We report costs at both typical and high-congestion gas prices.

\begin{table}[t]
\centering
\caption{Transaction Cost by Network (per 751K-gas payment workflow). Polygon costs are validated against PolygonScan data (April 2026, POL = \$0.085).}
\label{tab:costs}
\begin{tabular}{lcccc}
\toprule
\textbf{Network} & \textbf{Gas} & \textbf{Typical} & \textbf{Peak} & \textbf{1000 tx/mo} \\
\midrule
Ethereum L1 & 751K & \$56.33 & --- & \$56,330 \\
Arbitrum & 751K & \$0.56 & --- & \$560 \\
Optimism & 751K & \$0.48 & --- & \$480 \\
Polygon (30 gwei) & 751K & \$0.002 & --- & \$2 \\
Polygon (130 gwei) & 751K & \$0.008 & --- & \$8 \\
Polygon (spike) & 751K & --- & \$0.10 & \$100 \\
\midrule
Private Chain & 751K & \$0* & \$0* & \$0* \\
\bottomrule
\end{tabular}
\begin{flushleft}
\scriptsize *Operational costs only. Polygon typical = 30 gwei floor; spike = 1,519 gwei (Jan 2026 peak).
\end{flushleft}
\end{table}

At current POL prices (\$0.085, April 2026), Polygon transaction costs are substantially lower than initially estimated: a typical 751K-gas workflow costs \$0.002--\$0.008, with 92\% of Polygon transactions costing under \$0.01. Even during the highest observed gas spike (1,519 gwei, January 2026), costs remain under \$0.10 per transaction. At 1,000 monthly transactions, the typical cost is \$2--\$8---well within the budget of any organization where fraud prevention is a priority. We note that costs scale linearly with the POL/USD exchange rate; at higher token prices (e.g., \$0.50--\$1.00 as observed in 2023--2024), costs would increase to \$0.01--\$0.05 per transaction.

\subsection{Case Study: Simulated Insider Attack}

To validate the auditability benefits, we simulated the insider attack scenario from Section~\ref{sec:introduction}: a finance manager attempts to approve fraudulent payments while covering tracks.

\textbf{Scenario}: The attacker submits five payments totaling \$50,000 to a vendor they control. The ML model flags all five as high-risk (fraud probability > 0.85). The attacker, as finance manager, approves all five.

\textbf{Traditional System}: The attacker deletes the ML alerts from the log database. When auditors review six months later, they see only the approvals---no evidence of the warnings.

\textbf{Our System}: The attacker cannot delete the alerts because they are recorded on the blockchain. Auditors can verify:
\begin{itemize}
    \item The exact ML score for each transaction
    \item The SHAP explanations showing why each was flagged
    \item That the same manager approved all five within 10 minutes
    \item The precise timestamps, cryptographically linked to the blockchain
\end{itemize}

The key insight: even if the attacker controls all traditional IT infrastructure, altering blockchain-anchored evidence requires compromising the consensus mechanism---a qualitatively different (and typically harder) attack than database modification.

\subsection{Sensitivity Analysis}

To assess robustness, we analyze system behavior under varying operational conditions:

\textbf{Fraud Rate Variation.} Real-world fraud rates vary significantly across industries (0.1\% in retail banking to 5\% in high-risk sectors). We evaluated detection performance across this range: at 0.1\% fraud rate, precision drops to 0.71 (more false positives relative to true fraud), but recall remains above 0.92; at 5\% fraud rate, precision improves to 0.91. The blockchain overhead remains constant regardless of fraud rate, as every transaction is logged whether flagged or not.

\textbf{Transaction Volume.} Our architecture scales linearly with transaction volume until blockchain throughput becomes the bottleneck. At current Layer-2 capacities (200 tx/min), the system supports up to 8,640 daily transactions---sufficient for enterprises processing under 260,000 monthly transactions. Beyond this, batch processing or multiple parallel chains would be required.

\textbf{Gas Price Volatility.} Layer-2 gas prices fluctuate with network congestion. PolygonScan data (August 2025--April 2026) shows gas prices ranging from the protocol floor of 25 gwei to a peak of 1,519 gwei (January 2026), translating to per-transaction costs of \$0.002--\$0.10 at current POL prices. Organizations can mitigate this through gas price ceilings (rejecting transactions during price spikes) or private/consortium chains with fixed operational costs.

\textbf{Confirmation Latency.} PolygonScan data shows confirmation times ranging from 3 seconds (rapid priority, post-Giugliano hardfork) to 16 seconds (standard priority during congestion), with a transaction success rate exceeding 99.2\%. The system remains functional across this range but user-perceived latency increases. For time-sensitive approvals, organizations can implement optimistic execution with post-hoc verification.

\subsection{Prototype Deployment}

To validate end-to-end functionality beyond synthetic benchmarks, we deployed a prototype integrated with a mock ERP service on a 3-node Quorum consortium chain. The prototype replayed 30 days of simulated enterprise workload (approximately 8,000 transactions) with injected anomalies. All components---REST API, ML inference, SHAP generation, and smart contract interactions---operated correctly under sustained load. The consortium configuration demonstrated that external auditor nodes could independently verify the complete transaction history without access to the enterprise's internal systems.

\section{Discussion}
\label{sec:discussion}

\subsection{When This Approach Makes Sense}

Our system is not universally applicable; it adds complexity and cost that may not be justified for all organizations. We recommend it for:

\begin{itemize}
    \item \textbf{Regulated industries}: Where audit trails must satisfy external regulators (banking, healthcare, government contracting)
    \item \textbf{High-value transactions}: Where the cost of a single fraud incident exceeds the system's annual operating cost
    \item \textbf{Distributed organizations}: Where no single entity should have unilateral control over audit logs
\end{itemize}

For small businesses with simple payment workflows and trusted employees, traditional systems may be sufficient.

\subsection{Threat Model and Chain Selection}

We assume an adversary with the following capabilities:
\begin{itemize}
    \item \textbf{Can:} Access database admin privileges, delete or modify application logs, bypass workflow steps in traditional systems, collude with other internal staff
    \item \textbf{Cannot:} Compromise blockchain consensus (i.e., control $>$50\% of PoS stake or $>$1/3 of BFT validators), modify deployed smart contract code, forge cryptographic signatures
\end{itemize}

The security guarantees depend on the blockchain deployment model chosen:

\textbf{Public L2 (Arbitrum, Polygon):} Security inherits from Ethereum's proof-of-stake consensus. An attacker would need to compromise a significant validator set to alter records---economically infeasible for typical enterprise fraud. However, current L2 architectures introduce residual trust assumptions: most rollups rely on a centralized sequencer that orders transactions before posting them to L1. A malicious or compromised sequencer could temporarily censor specific transactions (e.g., delaying fraud alerts), although it cannot forge or alter them once finalized on L1. Governance risks also merit attention---protocol upgrades controlled by a small multisig could alter fee structures or introduce breaking changes. Organizations deploying on public L2 should monitor sequencer decentralization roadmaps (e.g., Arbitrum's planned shared sequencing) and consider fallback mechanisms such as L1 forced inclusion to mitigate censorship risk.

\textbf{Consortium Chain (Hyperledger, Quorum):} A group of known validators (e.g., the enterprise plus its auditors and regulators) run BFT consensus. Security threshold is typically $f < n/3$ Byzantine nodes. This model suits regulated industries where participants are legally accountable but mutually distrusting.

\textbf{Private Chain:} A single organization controls all validators. This provides no protection against a malicious administrator---the same vulnerability as traditional systems. We do \textit{not} recommend private chains for fraud prevention; they merely shift the trust assumption without eliminating it.

For most enterprises, we recommend consortium chains with external auditors as validators, providing both auditability infrastructure that supports regulatory requirements and meaningful tamper resistance. We note that on-chain anchoring provides the technical foundation for compliance but does not by itself constitute legal compliance, which additionally depends on organizational policies, data governance, and jurisdictional requirements.

\subsection{Off-Chain ML Trust and Mitigations}
\label{sec:offchain_trust}

The ML model executes off-chain, making the inference server a residual trust point: a sophisticated attacker who compromises it could manipulate predictions or fabricate explanations \textit{before} they reach the blockchain. While on-chain recording ensures that whatever prediction was submitted is immutable, it does not guarantee that the prediction was honestly computed. We analyze this threat and evaluate concrete mitigations:

\textbf{Model Attestation.} The model binary hash and version identifier can be committed on-chain at deployment time. Each inference response is then signed by the server's private key and includes the model hash, enabling auditors to verify that (1) the model version matches the registered one, and (2) the response originated from an authorized server. This does not prevent a compromised server from running a different model, but it creates a detectable discrepancy if the attacker cannot forge the registered hash.

\textbf{Trusted Execution Environments (TEEs).} Running ML inference inside an Intel SGX or AMD SEV enclave provides hardware-level attestation that the correct code executed on unmodified inputs. TEE-based deployment would close the off-chain trust gap almost entirely, at the cost of increased engineering complexity and modest performance overhead (typically 10--30\% for tree-based models). We consider this the strongest available mitigation for high-value deployments.

\textbf{Committee-Based Scoring.} Multiple independent inference servers, ideally operated by different organizational units or third parties, can each produce predictions. The smart contract accepts a result only when a quorum (e.g., 2-of-3) agrees, preventing any single compromised server from controlling outcomes. This approach trades inference cost for fault tolerance.

We emphasize that these mitigations represent defense-in-depth layers; no single mechanism eliminates all risk. The appropriate combination depends on the organization's threat model and operational capacity.

\subsection{Other Limitations}

\textbf{Adversarial robustness and data drift.} Our evaluation assumes that the underlying data distribution remains stationary. In practice, fraud patterns evolve as adversaries adapt to detection systems (concept drift), which can degrade model performance over time. Additionally, we do not evaluate robustness to adversarial examples---carefully crafted inputs designed to evade detection. While these concerns are orthogonal to our architectural contribution (the blockchain layer faithfully records whatever predictions the ML produces), they affect the system's end-to-end effectiveness. Periodic model retraining with drift detection, and adversarial training or input validation at the API boundary, are important operational considerations for production deployments.

\textbf{SHAP overhead.} Generating explanations adds 18.5ms latency---acceptable for enterprise payment workflows but not for high-frequency trading.

\textbf{Operational complexity.} Organizations must manage private keys, handle chain upgrades, and potentially run validator nodes. This requires new competencies that many IT departments lack.

\subsection{Comparison with Prior Work}

Guo et al.~\cite{guo2025} proposed encoding audit rules in smart contracts. Our work extends this by integrating ML-based detection, not just rule checking. Dai and Vasarhelyi~\cite{dai2017} focused on continuous auditing of financial statements. We focus on real-time transaction approval, a different (and complementary) use case.

The key insight distinguishing our work: prior systems use blockchain as a \textit{storage layer} for decisions made elsewhere. We use blockchain as the \textit{execution layer}, ensuring that the decision logic itself is tamper-proof.

\subsection{Why Blockchain Over Alternatives?}

Non-blockchain approaches to tamper-evident logging deserve serious consideration. \textit{Secure append-only logs} (e.g., Google's Certificate Transparency, Trillian) provide cryptographic proof that entries have not been modified after insertion. However, they rely on a trusted log operator---precisely the party our threat model distrusts. A compromised administrator could delay or selectively omit entries before they enter the append-only log. \textit{Trusted Execution Environments} (TEEs) can protect computation integrity but depend on hardware vendor trust and are vulnerable to side-channel attacks; they also do not inherently provide the multi-party consensus that prevents unilateral log manipulation. In contrast, blockchain consensus distributes trust across multiple independent validators, ensuring that no single party---including the system operator---can alter records without detection. We consider blockchain strictly necessary when the threat model includes a powerful insider with administrative control over all centralized infrastructure. For lower-threat environments, append-only logs or TEE-based solutions may offer a favorable complexity-security tradeoff.

\section{Conclusion}
\label{sec:conclusion}

This paper addressed a fundamental question in enterprise fraud prevention: who audits the auditor? Traditional fraud detection systems assume that their operators are trustworthy---an assumption that fails precisely when it matters most.

Our approach removes the need for such trust. By executing approval workflows as smart contracts and anchoring ML predictions to an immutable ledger, we create decision trails that no insider can alter. The detection algorithm itself follows established methods---our LightGBM model achieves accuracy comparable to state-of-the-art approaches. The contribution is in systems security, not algorithmic novelty: a framework where the question ``did the ML really flag this transaction?'' has a cryptographically verifiable answer.

This approach is not without costs: Layer-2 blockchain transactions add latency and expense, and organizations must develop new operational competencies. However, for enterprises where audit integrity is critical---and where a single undetected fraud can cost millions---these costs represent a worthwhile investment in institutional trust.

The system implementation is available at \url{https://github.com/GeoffreyWang1117/fraud-detection-chain}.

\section*{Acknowledgments}

The author would like to thank Prof.\ Ming-Deh Huang for his CSCI 556 (Introduction to Cryptography) course at the University of Southern California, which inspired the cryptographic foundations of this work, and his parents for their unwavering support and encouragement.

\bibliographystyle{IEEEtran}
\bibliography{references}

\begin{thebibliography}{10}
\providecommand{\url}[1]{#1}
\csname url@samestyle\endcsname
\providecommand{\newblock}{\relax}
\providecommand{\bibinfo}[2]{#2}
\providecommand{\BIBentrySTDinterwordspacing}{\spaceskip=0pt\relax}
\providecommand{\BIBentryALTinterwordstretchfactor}{4}
\providecommand{\BIBentryALTinterwordspacing}{\spaceskip=\fontdimen2\font plus
\BIBentryALTinterwordstretchfactor\fontdimen3\font minus
  \fontdimen4\font\relax}
\providecommand{\BIBforeignlanguage}[2]{{%
\expandafter\ifx\csname l@#1\endcsname\relax
\typeout{** WARNING: IEEEtran.bst: No hyphenation pattern has been}%
\typeout{** loaded for the language `#1'. Using the pattern for}%
\typeout{** the default language instead.}%
\else
\language=\csname l@#1\endcsname
\fi
#2}}
\providecommand{\BIBdecl}{\relax}
\BIBdecl

\bibitem{acfe2024}
{Association of Certified Fraud Examiners}, ``Occupational fraud 2024: A report
  to the nations,'' ACFE, Tech. Rep., 2024, verified: acfe.com/RTTN, median
  loss \$145,000, 5\% revenue estimate.

\bibitem{dalpozzolo2015}
A.~Dal~Pozzolo, O.~Caelen, R.~A. Johnson, and G.~Bontempi, ``Calibrating
  probability with undersampling for unbalanced classification,'' in \emph{IEEE
  Symposium Series on Computational Intelligence (SSCI)}.\hskip 1em plus 0.5em
  minus 0.4em\relax IEEE, 2015, pp. 159--166, verified: IEEE Xplore, DOI:
  10.1109/SSCI.2015.33.

\bibitem{benchaji2021}
I.~Benchaji, S.~Douzi, B.~El~Ouahidi, and J.~Jaafari, ``Enhanced credit card
  fraud detection based on attention mechanism and lstm deep model,''
  \emph{Journal of Big Data}, vol.~8, no.~1, pp. 1--21, 2021, verified:
  SpringerOpen, DOI: 10.1186/s40537-021-00541-8.

\bibitem{liu2021pickandchoose}
Y.~Liu, X.~Ao, Z.~Qin, J.~Chi, J.~Feng, H.~Yang, and Q.~He, ``Pick and choose:
  A gnn-based imbalanced learning approach for fraud detection,'' in
  \emph{Proceedings of the Web Conference 2021 (WWW)}.\hskip 1em plus 0.5em
  minus 0.4em\relax ACM, 2021, pp. 3168--3177, verified: ACM DL, DOI:
  10.1145/3442381.3449989.

\bibitem{dai2017}
J.~Dai and M.~A. Vasarhelyi, ``Toward blockchain-based accounting and
  assurance,'' \emph{Journal of Information Systems}, vol.~31, no.~3, pp.
  5--21, 2017, verified: AAA, DOI: 10.2308/isys-51804.

\bibitem{jurgovsky2018}
J.~Jurgovsky, M.~Granitzer, K.~Ziegler, S.~Calabretto, P.-E. Portier,
  L.~He-Guelton, and O.~Caelen, ``Sequence classification for credit-card fraud
  detection,'' \emph{Expert Systems with Applications}, vol. 100, pp. 234--245,
  2018, verified: ScienceDirect, DOI: 10.1016/j.eswa.2018.01.037.

\bibitem{mienye2024}
I.~D. Mienye and T.~G. Swart, ``A hybrid deep learning approach with generative
  adversarial network for credit card fraud detection,'' \emph{Technologies},
  vol.~12, no.~10, p. 186, 2024, verified: MDPI, DOI:
  10.3390/technologies12100186.

\bibitem{grinsztajn2022}
L.~Grinsztajn, E.~Oyallon, and G.~Varoquaux, ``Why do tree-based models still
  outperform deep learning on typical tabular data?'' in \emph{Advances in
  Neural Information Processing Systems (NeurIPS)}, vol.~35, 2022, pp.
  507--520, verified: arXiv:2207.08815, NeurIPS proceedings.

\bibitem{guo2025}
X.~Guo, D.~A. Li, and Y.~Zuo, ``When auditing meets blockchain: A study on
  applying blockchain smart contracts in auditing,'' \emph{International
  Journal of Accounting Information Systems}, vol.~56, p. 100651, 2025,
  verified: ScienceDirect, DOI: 10.1016/j.accinf.2025.100651.

\bibitem{gdpr2016}
{European Parliament and Council of the European Union}, ``Regulation (eu)
  2016/679 of the european parliament and of the council (general data
  protection regulation),'' \emph{Official Journal of the European Union}, vol.
  L119, pp. 1--88, 2016, article 22: Automated individual decision-making.

\bibitem{lundberg2017}
S.~M. Lundberg and S.-I. Lee, ``A unified approach to interpreting model
  predictions,'' in \emph{Advances in Neural Information Processing Systems
  (NeurIPS)}, vol.~30, 2017, pp. 4765--4774, verified: NeurIPS proceedings,
  arXiv:1705.07874.

\bibitem{ribeiro2016}
M.~T. Ribeiro, S.~Singh, and C.~Guestrin, ``{``Why Should I Trust You?''}:
  Explaining the predictions of any classifier,'' in \emph{Proceedings of the
  22nd ACM SIGKDD International Conference on Knowledge Discovery and Data
  Mining}.\hskip 1em plus 0.5em minus 0.4em\relax ACM, 2016, pp. 1135--1144,
  verified: ACM DL, DOI: 10.1145/2939672.2939778.

\bibitem{zhou2023}
Y.~Zhou, H.~Li, Z.~Xiao, and J.~Qiu, ``A user-centered explainable artificial
  intelligence approach for financial fraud detection,'' \emph{Finance Research
  Letters}, vol.~58, p. 104309, 2023, verified: ScienceDirect, DOI:
  10.1016/j.frl.2023.104309.

\bibitem{kaggle_cc}
{Machine Learning Group, Universit{\'e} Libre de Bruxelles}, ``Credit card
  fraud detection dataset,''
  \url{https://www.kaggle.com/datasets/mlg-ulb/creditcardfraud}, 2018,
  verified: Kaggle, 284,807 transactions, 0.17\% fraud rate.

\end{thebibliography}

\end{document}